

\catcode`@=11

\def\oneandahalfspace{\baselineskip=1.15\normalbaselineskip plus 1pt
\lineskip=2pt\lineskiplimit=1pt}

\def\nofirstpagenoten{\nopagenumbers\footline={\ifnum\pageno>1\tenrm
\hss\folio\hss\fi}}
\def\nofirstpagenotwelve{\nopagenumbers\footline={\ifnum\pageno>1\twelverm
\hss\folio\hss\fi}}
\def\leaderfill{\leaders\hbox to 1em{\hss.\hss}\hfill}
\def\ft#1#2{{\textstyle{{#1}\over{#2}}}}
\def\frac#1/#2{\leavevmode\kern.1em
\raise.5ex\hbox{\the\scriptfont0 #1}\kern-.1em/\kern-.15em
\lower.25ex\hbox{\the\scriptfont0 #2}}
\def\sfrac#1/#2{\leavevmode\kern.1em
\raise.5ex\hbox{\the\scriptscriptfont0 #1}\kern-.1em/\kern-.15em
\lower.25ex\hbox{\the\scriptscriptfont0 #2}}

\parindent=20pt
\def\narrow{\advance\leftskip by 40pt \advance\rightskip by 40pt}

\def\nonarrower{\advance\leftskip by -40pt\advance\rightskip by -40pt}

\def\boxit#1{\vbox{\hrule\hbox{\vrule\kern3pt
        \vbox{\kern3pt#1\kern3pt}\kern3pt\vrule}\hrule}}

\def\gtorder{\mathrel{\raise.3ex\hbox{$>$}\mkern-14mu
             \lower0.6ex\hbox{$\sim$}}}
\def\ltorder{\mathrel{\raise.3ex\hbox{$<$}|mkern-14mu
             \lower0.6ex\hbox{\sim$}}}
\def\dalemb#1#2{{\vbox{\hrule height .#2pt
        \hbox{\vrule width.#2pt height#1pt \kern#1pt
                \vrule width.#2pt}
        \hrule height.#2pt}}}
\def\square{\mathord{\dalemb{4.9}{5}\hbox{\hskip1pt}}}

\font\fourteentt=cmtt10 scaled \magstep2
\font\fourteenbf=cmbx12 scaled \magstep1
\font\fourteenrm=cmr12 scaled \magstep1
\font\fourteeni=cmmi12 scaled \magstep1
\font\fourteenssr=cmss12 scaled \magstep1
\font\fourteenmbi=cmmib10 scaled \magstep2
\font\fourteensy=cmsy10 scaled \magstep2
\font\fourteensl=cmsl12 scaled \magstep1
\font\fourteenex=cmex10 scaled \magstep2
\font\fourteenit=cmti12 scaled \magstep1
\font\twelvett=cmtt10 scaled \magstep1 \font\twelvebf=cmbx12
\font\twelverm=cmr12  \font\twelvei=cmmi12
\font\twelvessr=cmss12 \font\twelvembi=cmmib10 scaled \magstep1
\font\twelvesy=cmsy10 scaled \magstep1
\font\twelvesl=cmsl12 \font\twelveex=cmex10 scaled \magstep1
\font\twelveit=cmti12
\font\tenssr=cmss10 \font\tenmbi=cmmib10
 
 \font\ninebf=cmbx9
\font\ninerm=cmr9  \font\ninei=cmmi9
\font\ninesy=cmsy9 \font\ninessr=cmss9
\font\ninembi=cmmib10 scaled 900
\font\eightit=cmti8 \font\eightsl=cmsl8
\font\eighttt=cmtt8 \font\eightbf=cmbx8
\font\eightrm=cmr8  \font\eighti=cmmi8
\font\eightsy=cmsy8 \font\eightex=cmex10 scaled 800
\font\eightssr=cmss8 \font\eightmbi=cmmib10 scaled 800
 
\font\sevenbf=cmbx7 \font\sevenrm=cmr7 \font\seveni=cmmi7
\font\sevensy=cmsy7 
\font\sevenssr=cmss9 scaled 778 \font\sevenmbi=cmmib10 scaled 700
 
 \font\sixbf=cmbx7 scaled 875
\font\sixrm=cmr6  \font\sixi=cmmi6
\font\sixsy=cmsy6 \font\sixssr=cmss8 scaled 750
\font\sixmbi=cmmib10 scaled 600
\font\fivessr=cmss8 scaled 625  \font\fivembi=cmmib10 scaled 500

\newskip\ttglue
\newfam\ssrfam
\newfam\mbifam

\mathchardef\alpha="710B
\mathchardef\beta="710C
\mathchardef\gamma="710D
\mathchardef\delta="710E
\mathchardef\epsilon="710F
\mathchardef\zeta="7110
\mathchardef\eta="7111
\mathchardef\theta="7112
\mathchardef\iota="7113
\mathchardef\kappa="7114
\mathchardef\lambda="7115
\mathchardef\mu="7116
\mathchardef\nu="7117
\mathchardef\xi="7118
\mathchardef\pi="7119
\mathchardef\rho="711A
\mathchardef\sigma="711B
\mathchardef\tau="711C
\mathchardef\upsilon="711D
\mathchardef\phi="711E
\mathchardef\chi="711F
\mathchardef\psi="7120
\mathchardef\omega="7121
\mathchardef\varepsilon="7122
\mathchardef\vartheta="7123
\mathchardef\varpi="7124
\mathchardef\varrho="7125
\mathchardef\varsigma="7126
\mathchardef\varphi="7127
\mathchardef\partial="7140

\def\fourteenpoint{\def\rm{\fam0\fourteenrm}
\textfont0=\fourteenrm \scriptfont0=\tenrm \scriptscriptfont0=\sevenrm
\textfont1=\fourteeni \scriptfont1=\teni \scriptscriptfont1=\seveni
\textfont2=\fourteensy \scriptfont2=\tensy \scriptscriptfont2=\sevensy
\textfont3=\fourteenex \scriptfont3=\fourteenex \scriptscriptfont3=\fourteenex
\def\it{\fam\itfam\fourteenit} \textfont\itfam=\fourteenit
\def\sl{\fam\slfam\fourteensl} \textfont\slfam=\fourteensl
\def\bf{\fam\bffam\fourteenbf} \textfont\bffam=\fourteenbf
\scriptfont\bffam=\tenbf \scriptscriptfont\bffam=\sevenbf
\def\tt{\fam\ttfam\fourteentt} \textfont\ttfam=\fourteentt
\def\ssr{\fam\ssrfam\fourteenssr} \textfont\ssrfam=\fourteenssr
\scriptfont\ssrfam=\tenmbi \scriptscriptfont\ssrfam=\sevenmbi
\def\mbi{\fam\mbifam\fourteenmbi} \textfont\mbifam=\fourteenmbi
\scriptfont\mbifam=\tenmbi \scriptscriptfont\mbifam=\sevenmbi
\tt \ttglue=.5em plus .25em minus .15em
\normalbaselineskip=16pt
\bigskipamount=16pt plus5pt minus5pt
\medskipamount=8pt plus3pt minus3pt
\smallskipamount=4pt plus1pt minus1pt
\abovedisplayskip=16pt plus 4pt minus 12pt
\belowdisplayskip=16pt plus 4pt minus 12pt
\abovedisplayshortskip=0pt plus 4pt
\belowdisplayshortskip=9pt plus 4pt minus 6pt
\parskip=5pt plus 1.5pt
\twelvefoot
\setbox\strutbox=\hbox{\vrule height12pt depth5pt width0pt}
\let\sc=\tenrm
\let\big=\fourteenbig \normalbaselines\rm}
\def\fourteenbig#1{{\hbox{$\left#1\vbox to12pt{}\right.\n@space$}}
\def\square{\mathord{\dalemb{6.8}{7}\hbox{\hskip1pt}}}}

\def\twelvepoint{\def\rm{\fam0\twelverm}
\textfont0=\twelverm \scriptfont0=\ninerm \scriptscriptfont0=\sevenrm
\textfont1=\twelvei \scriptfont1=\ninei \scriptscriptfont1=\seveni
\textfont2=\twelvesy \scriptfont2=\ninesy \scriptscriptfont2=\sevensy
\textfont3=\twelveex \scriptfont3=\twelveex \scriptscriptfont3=\twelveex
\def\it{\fam\itfam\twelveit} \textfont\itfam=\twelveit
\def\sl{\fam\slfam\twelvesl} \textfont\slfam=\twelvesl
\def\bf{\fam\bffam\twelvebf} \textfont\bffam=\twelvebf
\scriptfont\bffam=\ninebf \scriptscriptfont\bffam=\sevenbf
\def\tt{\fam\ttfam\twelvett} \textfont\ttfam=\twelvett
\def\ssr{\fam\ssrfam\twelvessr} \textfont\ssrfam=\twelvessr
\scriptfont\ssrfam=\ninessr \scriptscriptfont\ssrfam=\sevenssr
\def\mbi{\fam\mbifam\twelvembi} \textfont\mbifam=\twelvembi
\scriptfont\mbifam=\ninembi \scriptscriptfont\mbifam=\sevenmbi
\tt \ttglue=.5em plus .25em minus .15em
\normalbaselineskip=14pt
\bigskipamount=14pt plus4pt minus4pt
\medskipamount=7pt plus2pt minus2pt
\abovedisplayskip=14pt plus 3pt minus 10pt
\belowdisplayskip=14pt plus 3pt minus 10pt
\abovedisplayshortskip=0pt plus 3pt
\belowdisplayshortskip=8pt plus 3pt minus 5pt
\parskip=3pt plus 1.5pt
\tenfoot
\setbox\strutbox=\hbox{\vrule height10pt depth4pt width0pt}
\let\sc=\ninerm
\let\big=\twelvebig \normalbaselines\rm}
\def\twelvebig#1{{\hbox{$\left#1\vbox to10pt{}\right.\n@space$}}
\def\square{\mathord{\dalemb{5.9}{6}\hbox{\hskip1pt}}}}

\def\tenpoint{\def\rm{\fam0\tenrm}
\textfont0=\tenrm \scriptfont0=\sevenrm \scriptscriptfont0=\fiverm
\textfont1=\teni \scriptfont1=\seveni \scriptscriptfont1=\fivei
\textfont2=\tensy \scriptfont2=\sevensy \scriptscriptfont2=\fivesy
\textfont3=\tenex \scriptfont3=\tenex \scriptscriptfont3=\tenex
\def\it{\fam\itfam\tenit} \textfont\itfam=\tenit
\def\sl{\fam\slfam\tensl} \textfont\slfam=\tensl
\def\bf{\fam\bffam\tenbf} \textfont\bffam=\tenbf
\scriptfont\bffam=\sevenbf \scriptscriptfont\bffam=\fivebf
\def\tt{\fam\ttfam\tentt} \textfont\ttfam=\tentt
\def\ssr{\fam\ssrfam\tenssr} \textfont\ssrfam=\tenssr
\scriptfont\ssrfam=\sevenssr \scriptscriptfont\ssrfam=\fivessr
\def\mbi{\fam\mbifam\tenmbi} \textfont\mbifam=\tenmbi
\scriptfont\mbifam=\sevenmbi \scriptscriptfont\mbifam=\fivembi
\tt \ttglue=.5em plus .25em minus .15em
\normalbaselineskip=12pt
\bigskipamount=12pt plus4pt minus4pt
\medskipamount=6pt plus2pt minus2pt
\abovedisplayskip=12pt plus 3pt minus 9pt
\belowdisplayskip=12pt plus 3pt minus 9pt
\abovedisplayshortskip=0pt plus 3pt
\belowdisplayshortskip=7pt plus 3pt minus 4pt
\parskip=0.0pt plus 1.0pt
\eightfoot
\setbox\strutbox=\hbox{\vrule height8.5pt depth3.5pt width0pt}
\let\sc=\eightrm
\let\big=\tenbig \normalbaselines\rm}
\def\tenbig#1{{\hbox{$\left#1\vbox to8.5pt{}\right.\n@space$}}
\def\square{\mathord{\dalemb{4.9}{5}\hbox{\hskip1pt}}}}

\def\eightpoint{\def\rm{\fam0\eightrm}
\textfont0=\eightrm \scriptfont0=\sixrm \scriptscriptfont0=\fiverm
\textfont1=\eighti \scriptfont1=\sixi \scriptscriptfont1=\fivei
\textfont2=\eightsy \scriptfont2=\sixsy \scriptscriptfont2=\fivesy
\textfont3=\eightex \scriptfont3=\eightex \scriptscriptfont3=\eightex
\def\it{\fam\itfam\eightit} \textfont\itfam=\eightit
\def\sl{\fam\slfam\eightsl} \textfont\slfam=\eightsl
\def\bf{\fam\bffam\eightbf} \textfont\bffam=\eightbf
\scriptfont\bffam=\sixbf \scriptscriptfont\bffam=\fivebf
\def\tt{\fam\ttfam\eighttt} \textfont\ttfam=\eighttt
\def\ssr{\fam\ssrfam\eightssr} \textfont\ssrfam=\eightssr
\scriptfont\ssrfam=\sixssr \scriptscriptfont\ssrfam=\fivessr
\def\mbi{\fam\mbifam\eightmbi} \textfont\mbifam=\eightmbi
\scriptfont\mbifam=\sixmbi \scriptscriptfont\mbifam=\fivembi
\tt \ttglue=.5em plus .25em minus .15em
\normalbaselineskip=9pt
\bigskipamount=9pt plus3pt minus3pt
\medskipamount=5pt plus2pt minus2pt
\abovedisplayskip=9pt plus 3pt minus 9pt
\belowdisplayskip=9pt plus 3pt minus 9pt
\abovedisplayshortskip=0pt plus 3pt
\belowdisplayshortskip=5pt plus 3pt minus 4pt
\parskip=0.0pt plus 1.0pt
\setbox\strutbox=\hbox{\vrule height8.5pt depth3.5pt width0pt}
\let\sc=\sixrm
\let\big=\eightbig \normalbaselines\rm}
\def\eightbig#1{{\hbox{$\left#1\vbox to6.5pt{}\right.\n@space$}}
\def\square{\mathord{\dalemb{3.9}{4}\hbox{\hskip1pt}}}}

\def\vfootnote#1{\insert\footins\bgroup\footsuite
    \interlinepenalty=\interfootnotelinepenalty
    \splittopskip=\ht\strutbox
    \splitmaxdepth=\dp\strutbox \floatingpenalty=20000
    \leftskip=0pt \rightskip=0pt \spaceskip=0pt \xspaceskip=0pt
    \textindent{#1}\footstrut\futurelet\next\fo@t}
\def\hangfootnote#1{\edef\@sf{\spacefactor\the\spacefactor}#1\@sf
    \insert\footins\bgroup\footsuite
    \let\par=\endgraf
    \interlinepenalty=\interfootnotelinepenalty
    \splittopskip=\ht\strutbox
    \splitmaxdepth=\dp\strutbox \floatingpenalty=20000
    \leftskip=0pt \rightskip=0pt \spaceskip=0pt \xspaceskip=0pt
    \smallskip\item{#1}\bgroup\strut\aftergroup\@foot\let\next}
\def\footsuite{}
\def\twelvefoot{\def\footsuite{\twelvepoint}}
\def\tenfoot{\def\footsuite{\tenpoint}}
\def\eightfoot{\def\footsuite{\eightpoint}}
\catcode`@=12
\twelvepoint
\oneandahalfspace
\nofirstpagenotwelve

\baselineskip=18pt
\def\h{{1\over 2}}

\def\p{\partial}
\def\lie{{\cal L}}
\def\tr{{\rm tr}}
\overfullrule=0pt

\line{\hfil CTP-TAMU-32/92}
\line{\hfil SISSA 182/92/EP}
\line{\hfil hepth@xxx/9210061}
\line{\hfil October 1992}
\vskip 2.4truecm
\centerline{\bf Symmetries of p--Branes}
\vskip 1truecm
\centerline{ R. Percacci}
\bigskip
\centerline{\it International  School for Advanced Studies,
via Beirut 4, 34014 Trieste, Italy}
\centerline{\it and Istituto Nazionale di Fisica Nucleare, Sezione di
Trieste.}
\bigskip
\centerline{and}
\bigskip
\centerline{E. Sezgin \footnote{$^\ast$}{\tenfoot Supported in part by the
U.S.\ National Science Foundation, under grant PHY-9106593.}}
\bigskip
\centerline{\it Center for Theoretical Physics, Texas A\&M University,}
\centerline{\it College Station, Texas 77843-4242, U.S.A.}
\vskip 1.4truecm
\centerline{\bf ABSTRACT}
\smallskip
\midinsert\narrower{
Using canonical methods, we study the invariance properties of a bosonic
$p$--brane propagating in a curved background locally diffeomorphic to $M\times
G$, where $M$ is spacetime and $G$ a group manifold. The action is that of a
gauged sigma model in $p+1$ dimensions coupled to a Yang--Mills field and a
$(p+1)$--form in $M$. We construct the generators of Yang-Mills and tensor
gauge transformations and exhibit the role of the $(p+1)$--form in cancelling
the potential Schwinger terms. We also discuss the Noether currents associated
with the global symmetries of the action and the question of the existence of
infinite dimensional  symmetry algebras, analogous to the Kac-Moody symmetry of
the string.}
\endinsert
\vfill\eject
\noindent {\bf 1. Introduction }
\medskip
The importance of symmetries in physical systems hardly needs to be
emphasized. In string theory, for example, the Virasoro and Kac-Moody
symmetries play a crucial role in solving and interpreting the theory.
In the case of higher $p$--branes, however, no analogues of these
symmetries are known. This motivates a systematic study of all invariances
of $p$--brane theory.

We choose to work with the bosonic $p$--brane theory discussed in [1].
It describes a $p$--brane  propagating in a curved background locally
diffeomorphic to $M\times G$, where $M$ is spacetime and $G$ is a group
manifold.  The $p$--brane is coupled to a metric, a Yang-Mills field and
a rank--$(p+1)$ antisymmetric tensor field on $M$.
Strictly speaking our conclusions apply only to this particular theory.
However we believe that our main conclusions apply also to fermionic
formulations of the $p$--brane. This is certainly true in the
case $p=1$, in which bosonization works. In the higher case the
bosonic and fermionic formulations are not equivalent. Instead, the
former can be thought of as a low--energy approximation of the latter.
We expect that symmetry aspects (such as anomaly cancellations)
are properly reflected in the low--energy theory.

If the Yang--Mills and tensor gauge field on $M$ are held fixed,
the model has world-volume diffeomorphisms as gauge invariances,
and Noether symmetries consisting of those transformations
which leave the background fields invariant. For generic background fields,
these will be just the group $G_L$ of left multiplications in $G$.
If the fields on $M$ are treated as dynamical, the theory is also
invariant under diffeomorphisms of $M$ and local $G_R$ transformations,
where $G_R$ is the group of right multiplications in $G$.
In this case one also has the Noether charges associated with global
$G_R$ transformations.

In this theory gauge invariance is achieved via a kind of
Green--Schwarz anomaly cancellation mechanism, with the variation of
the Wess--Zumino part of the action being compensated by the variation
of the term involving the antisymmetric tensor.
We will discuss in detail the counterpart of this phenomenon in the
canonical approach. We shall construct explicitly the generators
of Yang--Mills and tensor gauge transformations for arbitrary $p$--branes
and show that they form a closed algebra.
In the absence of the tensor gauge field
one would find an anomalous extension of the Yang--Mills algebra
of the Mickelsson--Faddeev type [2]. In the presence of the tensor gauge
field these anomalous extensions cancel; what remains can be
identified as a field-dependent tensor gauge transformation.
An alternative derivation of the gauge generators is given in [3].

It appears from all this that the true analogue of the Kac--Moody
symmetry of the string is the finite dimensional group $G_R$.
While the infinite dimensional target space gauge invariances
do not lead in the case of higher $p$--branes to true symmetry
algebras, they are nonetheless interesting for several reasons.
For example, the generators of these transformations
can be realized as functional differential operators on the
space of $p$--branes. These operators play a role in the study of
$p$--brane field theory.
Furthermore, they can be used to define functional
covariant derivatives which act on functionals of the $p$--brane.
These covariant derivatives become especially useful in the case of a
$\kappa$--invariant super $p$--brane theory [4], because their algebra
together with the principle of integrability along null super-planes
holds the key to the target space equations of motion [5,6].
Even in the absence of an action, they can be used to construct
the first class constraints of the theory, which may actually
{\it define} the theory.
We leave these applications for a future work.

The paper is organized as follows.
In section 2 we describe the model.
In section 3 we discuss the Noether
symmetries of the theory. In particular we find that
the algebra of Noether currents of global right and left
multiplications have an abelian extension.
In section 4 we treat the fields on $M$ as dynamical variables and derive the
currents which generate the gauge transformations.
The algebra of these currents is found to close,
as one would expect in a gauge invariant theory.
Throughout this paper we work with arbitrary odd $p$. Explicit formulae
for the cases $p=1,3,5$ are collected in an Appendix.
\medskip
\noindent {\bf 2. The Model}
\medskip
We will restrict our attention to the case of $p$--branes with $p$ odd.
The dynamical variables describing the $p$--brane are scalar fields
$x^\mu(\sigma)$, $y^m(\sigma)$ and a world-volume metric
$\gamma_{ij}(\sigma)$.
Here $\sigma^i\  (i=0,...,p)$ are the world-volume coordinates,
$x^\mu$, $\mu=0,...,d-1$ are coordinates on $M$ and
$y^m$, $m=1,..., dim\,G$ are coordinates on $G$.
The background fields are the metrics $g_{\mu\nu}(x)$ and $g_{mn}(y)$
on $M$ and $G$ respectively, a Yang-Mills field $A_\mu^a(x)$
and antisymmetric tensor fields $B_{\mu_1\ldots\mu_{p+1}}(x)$
and $b_{m_1\ldots m_{p+1}}(y)$.

The action for the $p$--brane propagating in curved background can
be written as [1]
$$
\eqalign{
S =&\int d^{p+1}\sigma {\cal L}\cr
=& \int d^{p+1}\sigma \sqrt{-\gamma}
\Bigl[-\ft12
\left(\gamma^{ij}\p_i x^\mu\p_j x^\nu g_{\mu\nu}
+\gamma^{ij}D_iy^m D_jy^n g_{mn}\right)+{p-1\over2}
+^\ast\! B+^\ast\! C-^\ast\! b\Bigr],\cr}\eqno(2.1)
$$
where $D_i y^m=\partial_i y^m-\partial_i x^\mu A_\mu^a L_a^m$.
In this paper we denote $L_a^m$ (resp. $R_a^m$) the left-invariant
(resp. right-invariant) Killing vectors on $G$, satisfying
$[L_a,L_b]=f^c{}_{ab}L_c$, $[R_a,R_b]=-f^c{}_{ab}R_c$,
$[L_a,R_b]=0$. The left-- (resp. right--) invariant Maurer-Cartan
forms are  $L_m=g^{-1}\p_m g=L_m^a T_a$, $R_m=\p_m g g^{-1}=R_m^a T_a$.
We follow the conventions of [1], namely the generators $T_a$
obey $[T_a,T_b]=f^c{}_{ab}T_c$ and the raising and lowering of algebra
indices is done with the invariant tensor
$d_{ab}=\tr\,T_aT_b$.
We will use form notation only for objects
defined on the world--volume. For example, we will denote
$A=\partial_i x^\mu A_\mu d\sigma^i$ and
$L=\partial_i y^m L_m d\sigma^i$ the pull-backs of the
connection on $M$ and of the Maurer-Cartan form on $G$.
Similarly, the forms $B$ and $b$ in (2.1) are given by
$$
\eqalign{
B=&\ {1\over(p+1)!}B_{\mu_1\ldots\mu_{p+1}}
\p_{i_1}x^{\mu_1}\cdots\p_{i_{p+1}}x^{\mu_{p+1}}
d\sigma^{i_1}\wedge\cdots\wedge d\sigma^{i_{p+1}}\ ,
\cr
b=&\ {1\over(p+1)!} b_{m_1\ldots m_{p+1}}
\p_{i_1}y^{m_1}\cdots\p_{i_{p+1}}y^{m_{p+1}}
d\sigma^{i_1}\wedge\cdots\wedge d\sigma^{i_{p+1}}\ .
\cr}\eqno(2.2)
$$
The tensor $B_{\mu_1\ldots\mu_{p+1}}$ is arbitrary,
while the tensor $b_{m_1\ldots m_{p+1}}$ is defined by the relation
$$
\p_{[m_1}b_{m_2\ldots m_{p+2}]}=
-k_pc_p(p+1)!{\rm tr}L_{[m_1}\ldots L_{m_{p+2}]}=
-k_pc_p(p+1)!{\rm tr}R_{[m_1}\ldots R_{m_{p+2}]}\ , \eqno(2.3)
$$
where $k_p$ and $c_p$ are normalization constants discussed in the Appendix.
The pulled--back version of (2.3) can be written
$$
db+\omega^0_{p+2}(L)=0\ , \eqno(2.4)
$$
where $\omega_{p+2}^0$ is a Chern-Simons form (see the Appendix).
The form $C$ in (2.1) is defined as follows. Let
$A_t=tA+(1-t)L$ and $F_t=dA_t+A_t^2=tF+t(t-1)(A-L)^2$.
Defining the operator
$$
\ell_t=dt(A^a-L^a){\partial\over\partial F_t^a}\ ,\eqno(2.5)
$$
we have
$$
C(A,L)=\int_0^1 \ell_t \omega_{p+2}^0(A_t,F_t)\ .\eqno(2.6)
$$
Explicit forms for the cases $p=1,3,5$ are given in the Appendix.
The Lagrangian ${\cal L}$ contains the duals of the forms $B$, $C$ and
$b$. The dual of a $p+1$--form $\omega$ is
$^\ast\omega={1\over (p+1)!}\varepsilon^{i_1\ldots i_{p+1}}
\omega_{i_1\ldots i_{p+1}}$.

The action is manifestly invariant under world--volume diffeomorphisms
and global $G_L$, which infinitesimally is given by
$\delta y^m=\epsilon^a R^m_a(y)$, where $\epsilon$ is a constant.
If the fields $A_\mu^a$ and $B_{\mu_1\ldots\mu_{p+1}}$ are treated
as independent variables, then the action is also invariant under
the tensor gauge transformations
$$
\delta_\Lambda B_{\mu_1\ldots\mu_{p+1}}=
(p+1)\partial_{[\mu_1}
\Lambda_{\mu_2\ldots\mu_{p+1}]} \eqno(2.7)
$$
and under the target space local $G_R$ transformations
$$
\eqalign{
\delta_\epsilon y^m=&\,\epsilon^a(x)L^m_a(y)\ ,\cr
\delta_\epsilon A_\mu^a=&\,\p_\mu\epsilon^a+f^a{}_{bc}A_\mu^b \epsilon^c\ ,\cr
\delta_\epsilon B_{\mu_1\ldots\mu_{p+1}}=&\,-(p+1)\p_{[\mu_1}\epsilon^a
\phi^a_{\mu_2\ldots\mu_{p+1}]}(A)\ ,\cr}\eqno(2.8)
$$
where $\phi^a(A)$ is a polynomial in $A_\mu$ and $F_{\mu\nu}$ which
is defined by equation (A.5).
The invariance is seen easily by noting that the
variations of the pulled-back fields are
$$
\eqalign{
\delta L=&\,d\epsilon+L\epsilon-\epsilon L\ ,\cr
\delta A=&\,d\epsilon+A\epsilon-\epsilon A\ ,\cr
\delta B=& -\omega^1_{p+1}(A,\epsilon)+d\Lambda\ ,\cr}
\eqno(2.9)
$$
where $\omega^1_{p+1}$ is defined by (A.4).
Under the transformations (2.9), we find (up to surface terms)
from equation (2.4)
$\delta b=-\omega^1_{p+1}(L,\epsilon)$, while
$\delta C=\omega^1_{p+1}(A,\epsilon)-\omega^1_{p+1}(L,\epsilon)$,
so the action (2.1) is gauge invariant.

The algebra of the gauge transformations (2.8) is derived by
using the Wess--Zumino consistency condition (formula (A.7) in the
Appendix) and reads
$$
\eqalignno{
[\delta_{\epsilon_1},\delta_{\epsilon_2}]=&\
\delta_{[\epsilon_1,\epsilon_2]}+\delta_{\Lambda=\omega^2_p}\ , &(2.10a)\cr
[\delta_\epsilon,\delta_\Lambda]=&\ 0\ ,&(2.10b)\cr
[\delta_{\Lambda_1},\delta_{\Lambda_2}]=&\ 0\ ,&(2.10c)\cr}
$$
where $\omega^2_p(A,\epsilon_1,\epsilon_2)$ is the 2-cocycle defined
in (A.7).

Notice that $\int(C-b)$ is the Wess--Zumino action, whose infinitesimal
gauge variation is the consistent anomaly
$\omega^1_{p+1}(A,\epsilon)$. The difference between the usual
gauged Wess--Zumino model and the present theory resides in the fact
that here the gauge field $A$ is not a fundamental variable but rather
a composite field, {\it i.e.} a fixed functional of the scalar fields $x$.
Therefore the anomaly that
would be present in the absence of the $B$ field is a so--called
sigma model anomaly.

If we treat also the spacetime metric as an independent
variable, then the theory is manifestly invariant under target space
diffeomorphisms. Unlike the case of the gauge symmetry discussed
above, there is no subtle anomaly cancellation involved here, so
we shall not discuss this invariance further in this paper.
\medskip
\noindent{\bf 3. The Noether Symmetries}
\medskip
In this section we will treat the fields $A_\mu^a$,
$B_{\mu_1\ldots\mu_{p+1}}$ and $g_{\mu\nu}$ as fixed backgrounds,
as is customary when the $p$--brane is treated as a fundamental theory.
Since the background fields are not to be varied, the only
invariances of the theory are the world--volume diffeomorphisms
and the global transformations which leave the background fields
invariant. In the case of the string there is in addition Weyl
invariance. In this case the Virasoro and the Kac--Moody groups
are infinite dimensional global symmetries with associated Noether
charges. No such infinite dimensional symmetries are known in the case
of higher $p$--branes with the usual action.
We shall illustrate this fact below in the case of Kac--Moody symmetry.

For general $p$--branes the world--volume diffeomorphisms will have
vanishing Noether charges, as one would expect of a gauge invariance.
We will not discuss them any further in this paper.

We will only be interested in variations of the form
$\delta x^\mu=0$, $\delta y^m=v^m(y)$.
Then the invariance condition reads
$$
\eqalignno{
0=\delta S= \int d^{p+1}\sigma\,& \Biggl[
\sqrt{-\gamma}\gamma^{ij} D_j y^m L_m^a
\partial_jy^n \lie_v L_n^a \cr
&-{1\over(p+1)!}\varepsilon^{i_1\ldots i_{p+1}}
\partial_{i_1}y^{m_1}\cdots \partial_{i_{p+1}}y^{m_{p+1}}
{\cal L}_vb_{m_1\ldots m_{p+1}}\cr
&+{\delta \int^\ast\! C\over\delta L_i^a(\sigma)}
\partial_i y^m{\cal L}_v L_m^a \Biggr]
\ .  &(3.1) \cr}
$$
This can be satisfied by choosing $v$ so that ${\cal L}_v L_m^a=0$ and
$$
{\cal L}_vb_{m_1\ldots m_{p+1}}=
(p+1)\partial_{[m_1}\lambda_{m_2\ldots m_{p+1}]} \eqno(3.2)
$$
for some $p$-form $\lambda$ on $G$.
These conditions are satisfied by the global left group action
$v^m=\epsilon^a R_a^m$, where $\epsilon$ is {\it constant}.
The Noether current corresponding to this transformation is
$$
J^i_{Ra}=R_a^m\left(\sqrt{-\gamma}\gamma^{ij}D_jy^m+
L_m^b{\partial ^\ast C\over\partial L^b_i}\right)
+{1\over p!}\varepsilon^{ii_1\ldots i_p}\partial_{i_1}y^{m_1}\cdots
\partial_{i_p}y^{m_p}\lambda^a_{m_1\ldots m_p}\ ,\eqno(3.3)
$$
where the $p$--form $\lambda^a_p$ is defined by
$\lambda_p=\epsilon_a \lambda^a_p$. From (2.3) and (3.2) one gets
$$
\lambda^a_{m_1\ldots m_p}=
R_a^m b_{mm_1\ldots m_p}
-k_pc_p{(p+2)!\over p+1}\tr \,T^a R_{m_1}\cdots R_{m_p}\ . \eqno(3.4)
$$
The momenta canonically conjugate to $y^m$ are
$$
p_m={\partial{\cal L}\over\partial\partial_0 y_m}=
\sqrt{-\gamma}\gamma^{0j}D_jy^ng_{mn}-
{1\over p!}\varepsilon^{r_1\ldots r_p}
\p_{r_1}y^{m_1}\cdots\p_{r_p}y^{m_p}b_{mm_1\ldots m_p}
+L_m^a{\partial ^\ast C\over\partial L_0^a}\ ,\eqno(3.5)
$$
where $g_{mn}=L_m^aL_n^bd_{ab}$ is the invariant metric on $G$ and
$r=1\ldots p$ refer to spacelike directions of the world--volume.
Then, for $i=0$ the charge density can be rewritten as
$$
J_{Ra}=R_a^m\!\left(p_m+
{1\over p!}\varepsilon^{r_1\ldots r_p}\partial_{r_1}y^{m_1}\cdots
 \partial_{r_p}y^{m_p}\!\left(b_{mm_1\ldots m_p}
-k_pc_p{(p+2)!\over p+1}\tr\,R_mR_{m_1}\ldots R_{m_p}\right)\!\right).
\eqno(3.6)
$$
The algebra of the charge densities is
$$
\eqalign{
\{J_{Ra}(\sigma),&J_{Rb}(\sigma^\prime)\}=
-f_{abc}J_{Rc}(\sigma)\delta^{(p)}(\sigma,\sigma^\prime) \cr
&+2 k_p c_p(p+2)\varepsilon^{r_1\ldots r_{p}}
\tr \left(\{T_a,T_b\}R_{r_1}\cdots R_{r_{p-1}}\right)
\partial_{r_p}\delta^{(p)}(\sigma,\sigma^\prime)\ .\cr}\eqno(3.7)
$$
where $R_r=\partial_{r}y^{m}R_m$.
Note that the extension integrates to zero on a spacelike surface,
so the algebra of
the Noether charges $Q_{Ra}=\int d^p\sigma J_{Ra}(\sigma)$ is
(anti-)isomorphic to the Lie algebra of $G$.

For strings ($p$=1) the Noether symmetry group is much larger. If we set
$v^m=\epsilon^a(\sigma)R_a^m$, condition (3.1) reduces to
$$
\left(\sqrt{-\gamma}\gamma^{ij}-\varepsilon^{ij}\right)D_j\epsilon^a=0
\ .
\eqno(3.8)
$$
For $A_\mu=0$ this equation admits all functions of $\sigma^0+\sigma^1$
as solutions.
These form an infinite dimensional Kac-Moody algebra with a central extension.
It is a Noether symmetry because the corresponding Noether charges are
nonvanishing.
One could ask whether a similar algebra exists also for higher $p$--branes.
Restricting our attention for simplicity to the case $A_\mu=0$, the
invariance condition (3.1) reduces to
$$
\int d^{p+1}\sigma
\left(\sqrt{-\gamma}\gamma^{ij}\delta^{ab}+k_p c_p{(p+2)!\over p+1}
\varepsilon^{ijk_1\ldots k_{p-1}}\tr\, T^aT^bL_{k_1}\cdots
L_{k_{p-1}}\right)\partial_j\epsilon^a=0\ .\eqno(3.9)
$$
As opposed to the case of eq. (3.8),
the coefficient matrix in brackets is now a
functional of the fields $y^m$. For generic fields this matrix is
nondegenerate and therefore the only solution is $\epsilon^a$ constant.
Consequently, the corresponding Noether charges are just the usual
global Yang--Mills charges.

Finally we note that in the absence of gauge fields the theory
would also have a global right $G$ invariance, with associated
Noether charge density
$$
J_{La}=L_a^m\left(p_m+
{1\over p!}\varepsilon^{r_1\ldots r_p}\partial_{r_1}y^{m_1}\cdots
 \partial_{r_p}y^{m_p}\left(b_{mm_1\ldots m_p}+
k_p c_p{(p+2)!\over p+1}\tr\,L_mL_{m_1}\ldots L_{m_p}\right)\right)\
.\eqno(3.10)
$$
Note that using (A.5) and (A.12b) the last term can be rewritten
as $^\ast\phi^a_p(L)$. The algebra of these currents is
$$
\eqalign{
\{J_{La}(\sigma),&J_{Lb}(\sigma^\prime)\}=
f_{abc}J_{Lc}(\sigma)\delta^{(p)}(\sigma,\sigma^\prime) \cr
&-2k_p c_p(p+2)\varepsilon^{r_1\ldots r_{p}}
\tr \left(\{T_a,T_b\}L_{r_1}\cdots L_{r_{p-1}}\right)
\partial_{r_p}\delta^{(p)}(\sigma,\sigma^\prime)\ .\cr}\eqno(3.11)
$$
For later use we observe that multiplying this formula by
$\epsilon_1(\sigma)\epsilon_2(\sigma')$ (not to be confused with the constant
symmetry transformation parameters), integrating over
$\sigma$, $\sigma'$ and making use of equation (A.12c),
we can write
$$
\{J_{L\epsilon_1},J_{L\epsilon_2}\}=
J_{L[\epsilon_1,\epsilon_2]}+\omega^2_p(L,\epsilon_1,\epsilon_2)\ .
\eqno(3.12)
$$
In the case $p=3$ this algebra was derived using canonical methods
in [7] and from a different point of view in [8].
\medskip
\noindent{\bf 4. Target Space Gauge Invariance}
\medskip
As mentioned above, if we allow the background fields
to be transformed in a suitable way, the theory is invariant under
target-space dependent gauge transformations. This invariance arises
through a cancellation of anomalous terms. In fact, if one drops the
field $x$ and treats $A$ as a fundamental (rather than composite)
gauge field, then the action (2.1) (without the term $^\ast B$)
describes a gauged sigma model with Wess--Zumino term.
This model is well-known not to be gauge invariant and moreover its anomaly
cannot be cancelled by the introduction of a fundamental $B$ field
(the field equation for $B$ would be inconsistent). However, in our
model $A$ is a composite gauge field and the resulting anomaly is
known as a sigma model anomaly. In this case a composite $B$
field can be meaningfully employed to cancel the anomaly {\it via}
the Green--Schwarz mechanism (for strings, this was illustrated in [9]).

In this section, we are going to discuss this anomaly-cancellation
mechanism at the Hamiltonian level. As is well known, anomalies appear
in the Hamiltonian formulation as Schwinger terms in the algebra of
the currents which couple to the gauge fields.
A convenient way of dealing with this problem is to treat the
fields $A_\mu^a$ and $B_{\mu_1\ldots\mu_{p+1}}$
as dynamical variables.
In this case the action which describes the dynamics of the extended objects
coupled to these fields can be written as
$$
S=\int d^{p+1}\sigma\int d^dx\delta^d(x,x(\sigma)){\cal L}\
,\eqno(4.1)
$$
where ${\cal L}$ is defined as in (2.1), with
$A_\mu$ and $B_{\mu_1\ldots\mu_{p+1}}$ now regarded as functions of
$x$ rather than $x(\sigma)$.
At this point we could also add to ${\cal L}$ independent kinetic
terms for $A$ and $B$, but we shall not do so here.
Note that we could have introduced also a factor $\delta(y,y(\sigma))$
and an integration over $y$. However, since the fields
$A_\mu$ and $B_{\mu_1\ldots\mu_{p+1}}$ are $y$-independent,
this procedure is unnecessary. Therefore it is always understood that
$y=y(\sigma)$.

The equations of motion for the spacetime fields are
$$
\eqalignno{
0=&j^a_\mu(x)=\ {\delta S\over\delta A_\mu^a(x)}=
-\int d^{p+1}\sigma \delta^d(x,x(\sigma))
\sqrt{\gamma}\gamma^{ij}\partial_ix^\mu D_i y^m L_m^a
+{\delta \int ^\ast\! C\over\delta A_\mu^a(x)}\ , &(4.2a)\cr
0=&\ j^{\mu_1\ldots\mu_{p+1}}(x)=\,
{\delta S\over\delta B_{\mu_1\ldots\mu_{p+1}}(x)}=
\varepsilon^{i_1\ldots i_{p+1}}\partial_{i_1}x^{\mu_1}\cdots\
\partial_{i_{p+1}}x^{\mu_{p+1}}\ .&(4.2b)\cr}
$$
For simplicity of notation from now on the symbol $\int$ of
indefinite integration will stand for
$\int d^dx\int d^{p+1}\sigma\delta^d(x,x(\sigma))$.
Notice that owing to the absence of kinetic terms, the equations (4.2)
do not contain second time derivatives of the fields and are therefore
equations of constraint.

In the Hamiltonian formulation of this theory one has primary constraints
$$
\eqalignno{
P^\mu_a-{\partial ^\ast C\over \partial\partial_0 A_\mu^a}=&\,0, &(4.3a)\cr
P^{\mu_1\ldots\mu_{p+1}}=&\,0, &(4.3b)\cr}
$$
where $P_a^\mu$ and $P^{\mu_1\ldots\mu_{p+1}}$
are the momenta canonically conjugate to
$A_\mu^a$ and  $B_{{\mu_1\ldots\mu_{p+1}}}$ respectively.
These primary constraints arise from the fact that kinetic terms have
not been included for $A_\mu^a$ and  $B_{{\mu_1\ldots\mu_{p+1}}}$.
In fact, demanding that the primary constraints have vanishing Poisson brackets
with the Hamiltonian, one finds the secondary constraints which are
equivalent to the field equations (4.2).
Since the theory is gauge invariant, there must exist linear
combinations of these constraints which are first class and generate
the gauge transformations.

We choose the gauge for world-volume diffeomorphisms such that $x^0=\sigma^0$.
Then from (4.2), using (3.5), we obtain
$$
\eqalignno{
j^0_a(x)=&\int d^{p+1}\sigma \delta^d(x,x(\sigma))
L_a^m\left(p_m+
{1\over p!}\varepsilon^{r_1\ldots r_p}\partial_{r_1}y^{m_1}\cdots
\partial_{r_p}y^{m_p}b_{mm_1\ldots m_p}\right)\cr
&\qquad\qquad\qquad\qquad
-\left(
{\delta \int ^\ast\! C\over\delta A_0^a(\sigma)}+
{\delta \int ^\ast\! C\over\delta L_0^a(\sigma)}\right)\ ,
&(4.4)\cr
j^{0\mu_1\ldots\mu_p}(x)=&\int d^{p+1}\sigma \delta^d(x,x(\sigma))
\varepsilon^{r_1\ldots r_p}\partial_{r_1}x^{\mu_1}\cdots
\partial_{r_p}x^{\mu_{p}}\ . &(4.5)\cr}
$$
In this formula and in the rest of the paper, we use
world-volume pullbacks $A_0^a=\partial_0x^\mu
A_\mu^a$ and $L_0^a=\p_0y^mL_m^a$. Note that because $dL=-L^2$
we can assume that no derivatives of $L$ appear in $C$, and
therefore
${\delta ^\ast\! C(\sigma ')\over\delta L_0^a(\sigma)}=
{\p ^\ast\! C\over\p L_0^a}\delta(\sigma,\sigma')$.

We find that the first class constraints generating
tensor gauge transformations are
$$
G_\Lambda=\int \left(
(p+1)\partial_{\mu_1}\Lambda_{\mu_2\ldots\mu_{p+1}}P^{\mu_1\ldots\mu_{p+1}}-
{1\over p!}\Lambda_{\mu_1\ldots\mu_p}j^{0\mu_1\ldots\mu_p}\right)\eqno(4.6)
$$
and those generating Yang--Mills gauge transformations are
$$
\eqalign{G_\epsilon=&\int
\Big[D_\mu\epsilon^a
\left(P^\mu_a-{\p ^\ast C\over\p\p_0 A_\mu^a}\right)
+\epsilon^a j^0_a\cr
& -(p+1)(\partial_{\mu_1}\epsilon^a) \phi^a_{\mu_2\ldots\mu_{p+1}}(A)
P^{\mu_1\ldots\mu_{p+1}}
+{1\over p!}\epsilon^a \phi^a_{\mu_1\ldots\mu_p}(A)j^{0\mu_1\ldots\mu_p}\Big].}
\eqno(4.7)
$$
where $\phi_p$ is defined in (A.5).
These equations can be understood as follows.
The coefficients of the momenta $P_a^\mu$, $p_m$ and
$P^{\mu_1\ldots\mu_{p+1}}$ are fixed by the requirement that
the Poisson brackets of $G_\epsilon$ with the fields
$A_\mu^a$, $y^m$ and $B_{\mu_1\ldots\mu_{p+1}}$ yield the
gauge transformations (2.9). Since the momenta appear linearly in
the constraints, this fixes the coefficients of the first three
terms. The coefficient of the last term is fixed by the
requirement that the generators form a closed algebra.

To prove that this happens we first simplify the form of the
generator $G_\epsilon$. We observe that the second term
in round brackets in (4.7) and the third term in
(4.4) combine as follows:
$$
\int \left(- D_\mu\epsilon^a {\p ^\ast C\over\p\p_0 A_\mu^a}+
\epsilon^a{\delta ^\ast C\over\delta A_0^a}\right)
=\int d^dx \epsilon^a{\p ^\ast C\over\p A_0^a}\ . \eqno(4.8)
$$
The partial derivative on the r.h.s. of this formula means that
$C$ should be written in terms of $A$ and $F$ and varied only with
respect to $A$.
Thus, the first line in (4.7) can be written as
$$
G_\epsilon^{(1)}=\int \left[D_\mu\epsilon^a P^\mu_a
+\epsilon^a L_a^m\left(p_m+
{1\over p!}\varepsilon^{r_1\ldots r_p}\partial_{r_1}y^{m_1}\cdots
\partial_{r_p}y^{m_p}b_{mm_1\ldots m_p}\right)-\epsilon^a V_a\right]\ ,
\eqno(4.9)
$$
where
$$
V_\epsilon=\epsilon^a V_a=
\epsilon^a\left({\p ^\ast C\over \p A_0^a}
+{\p ^\ast C\over \p L_0^a}\right)\ .
\eqno(4.10)
$$
To compute $V$, let us introduce a (graded) derivation $\ell_\epsilon$,
defined by $\ell_\epsilon A=\ell_\epsilon L=\epsilon$, $\ell_\epsilon F=0$.
Then we can write $V_\epsilon=\ell_\epsilon C$. We now use the formula (2.6)
for $C$. We observe that $\ell_\epsilon$ anticommutes with $\ell_t$
and that acting on $\omega^0_{p+2}(A_t,F_t)$ it coincides with the
operator $\ell_\lambda$ defined in the Appendix of [10] where it is also
shown that $\ell_\lambda \omega^0_{p+2}=\hat\omega^1_{p+1}$
(see (A.9)). Using these results we obtain
$$
V_\epsilon=-\int_0^1 \ell_t\ell_\epsilon \omega^0_{p+2}(A_t,F_t)=
\int_0^1 \ell_t\tr\ \epsilon d\phi_p(A_t,F_t)\ .
$$
We now apply to $\phi_p$ the homotopy formula [10]
$$
d_t=dt{d\over dt}=(\ell_t d-d\ell_t)\eqno(4.11)
$$
with $\ell_t$ given by equation (2.5).  We find that
$$
V_a=\phi_p^a(A)-\phi_p^a(L)+d\chi^a_{p-1}\ ,\eqno(4.12)
$$
where
$$
\chi^a_{p-1}(A,L)=\int_0^1 \ell_t\phi_p^a(A_t,F_t)\ .\eqno(4.13)
$$
The explicit expressions for $\chi^a_{p-1}$ in the cases $p=1,3,5$
is given in (A.13). Substituting (4.12) in (4.9), the second term in $V_a$,
which from (A.12b) is equal to $c_p(p+2)\tr\,T_aL^p$, combines with the purely
$y$-dependent terms in $G_\epsilon$ to yield the generator of the global right
multiplications given in (3.10). We can thus write
$G_\epsilon=G_\epsilon^{(1)}+G_\epsilon^{(2)}$, with
$$
\eqalignno{
G_\epsilon^{(1)}=&\int \left[D_\mu\epsilon^a P^\mu_a+
\epsilon^a(J_L^a-d\chi_{p-1}^a(A,L)-\phi_p^a(A))\right]\ , &(4.15a)\cr
G_\epsilon^{(2)}=&\int \left[
-(p+1)\partial_{\mu_1}\epsilon^a \phi^a_{\mu_2\ldots\mu_{p+1}}(A)
P^{\mu_1\ldots\mu_{p+1}}
+\epsilon^a \phi_p^a(A)\right]\ .&(4.15b)\cr}
$$
For the reader's convenience we recall that $J_L$, $\chi_a$ and $\phi$
are defined in (3.10), (4.13) and (A.5) respectively.
Note that the terms of the form $\tr\epsilon\phi$ cancel in
$G_\epsilon$ and one gets
$$
G_\epsilon= \Big[ D_\mu\epsilon^aP^\mu_a
-(p+1)\partial_{\mu_1}\epsilon^a \phi^a_{\mu_2\ldots\mu_{p+1}}
P^{\mu_1\ldots\mu_{p+1}}
+\epsilon^a(J_L^a-d\chi_{p-1}^a)\big]\ . \eqno(4.16)
$$
 Even though for the purpose of computing the
algebra it would be more efficient to make use of this
simplification, it is instructive to keep $G_\epsilon^{(1)}$
and $G_\epsilon^{(2)}$ separate.
In fact, $G_\epsilon^{(1)}$ is identical to the
Gauss law operator of the gauged Wess-Zumino-Witten model,
except for the fact that the gauge field $A$ is now
composite. On the other hand $G_\epsilon^{(2)}$ is a linear
combination of the constraints which follow from the existence
of the field $B$, and has no analogue in the gauged Wess-Zumino-Witten
model.

We will now compute separately the Poisson brackets of $G_\epsilon^{(1)}$
and $G_\epsilon^{(2)}$. Using (3.12) we have
$$
\eqalign{
\{G^{(1)}_{\epsilon_1},G^{(1)}_{\epsilon_2}\}=&
\int \Bigl[D_\mu[\epsilon_1,\epsilon_2]^a P^\mu_a
+[\epsilon_1,\epsilon_2]^aJ_{La}+\omega^2_p(L,\epsilon_1,\epsilon_2)\cr
&-\delta_{\epsilon_1}(\epsilon_2^ad\chi_{p-1}^a)
+\delta_{\epsilon_2}(\epsilon_1^ad\chi_{p-1}^a)
-\delta_{\epsilon_1}(\epsilon_2^a\phi_p^a)
+\delta_{\epsilon_2}(\epsilon_1^a\phi_p^a)
\Bigr]\cr}\ .\eqno(4.17)
$$
The last four terms in this formula arise from the
Poisson brackets of the first two terms with the last two
terms in (4.15a).
Applying the homotopy formula to $\omega_p^2$ one gets the identity
$$
\delta_{\epsilon_1}\int \tr\,\epsilon_2d\chi_{p-1}-
\delta_{\epsilon_2}\int \tr\,\epsilon_1d\chi_{p-1}
=\int \left(\tr\,[\epsilon_1,\epsilon_2]d\chi_{p-1}
-\omega_p^2(A,\epsilon_1,\epsilon_2)
+\omega_p^2(L,\epsilon_1,\epsilon_2)\right)\ . \eqno(4.18)
$$
On the other hand subtracting algebraically equation (A.10) from (A.7)
we find that
$$
\delta_{\epsilon_1}\int \tr\,\epsilon_2\phi_p-
\delta_{\epsilon_2}\int \tr\,\epsilon_1\phi_p
=\int \left(\tr\,[\epsilon_1,\epsilon_2]\phi_p
+\omega_p^2(A,\epsilon_1,\epsilon_2)-
\hat\omega_p^2(A,\epsilon_1,\epsilon_2)\right)\ . \eqno(4.19)
$$
Substituting in (4.16) we find that the operators $G_\epsilon^{(1)}$
satisfy the algebra
$$
\{G^{(1)}_{\epsilon_1},G^{(1)}_{\epsilon_2}\}=
G^{(1)}_{[\epsilon_1,\epsilon_2]}
+\int \hat\omega_p^2(A,\epsilon_1,\epsilon_2)\ . \eqno(4.20)
$$
This agrees with the explicit calculations in the Wess-Zumino-Witten
model in two and four dimensions [11]. (It is interesting to note that
if one adds to $G_\epsilon^{(1)}$ the term $\int \phi_p^a(A)$, then
by using (4.19), one finds that the anomalous extension in (4.20) gets
replaced by $\omega_p^2(A,\epsilon_1,\epsilon_2)$). Next, using the
Wess--Zumino consistency condition (A.7) for $\omega^1_{p+1}$, we get
$$
\eqalignno{
\{G^{(1)}_{\epsilon_1},G^{(2)}_{\epsilon_2}\}-
\{G^{(1)}_{\epsilon_2},G^{(2)}_{\epsilon_1}\}=&
\int \Bigl[\left(-\omega^1_{p+1}(A_\mu,[\epsilon_1,\epsilon_2])
-d\omega_p^2(A_\mu,\epsilon_1,\epsilon_2)\right)_{\mu_1\ldots\mu_{p+1}}
P^{\mu_1\ldots\mu_{p+1}}\cr
&\qquad\qquad\qquad\qquad
+\delta_{\epsilon_1}(\epsilon_2^a\phi_p^a)
-\delta_{\epsilon_2}(\epsilon_1^a\phi_p^a)\Bigr]\cr
=&\ G^{(2)}_{[\epsilon_1,\epsilon_2]}-
G_{\Lambda(A,\epsilon_1,\epsilon_2)}
-\int\hat\omega_p^2(A,\epsilon_1,\epsilon_2) &(4.21)\cr}
$$
where $G_\Lambda$ is the generator of a field--dependent tensor gauge
transformation with parameter
$\Lambda(A,\epsilon_1,\epsilon_2)=\omega^2_p(A,\epsilon_1,\epsilon_2)$.
Collecting (4.20), (4.21) and observing that
$\{G_{\epsilon_1}^{(2)},G_{\epsilon_2}^{(2)}\}=0$, we
see that the anomalous extensions cancels and we remain with
$$
\{G_{\epsilon_1},G_{\epsilon_2}\}=G_{[\epsilon_1,\epsilon_2]}
-G_{\Lambda(A,\epsilon_1,\epsilon_2)}\ ,\eqno(4.22)
$$
with $G_\epsilon$ and $G_\Lambda$ given in (4.16) and (4.6), respectively.
Since evidently $G_\Lambda$ has vanishing Poisson brackets
with all other generators, the algebra of the
generators of Yang--Mills and tensor gauge
transformations closes with field--dependent structure constants.
It is isomorphic to the algebra given in (2.10).
It is important to stress the difference between the significance
of (4.20) and (4.22).
The former is referred to in the literature on anomalies as
a Mickelsson--Faddeev algebra; the second term on its right
hand side cannot be identified with any generator of the algebra
and therefore gives rise to an anomalous extension.
In the latter, the second term on the right hand side is an
already existing generator and therefore should not be regarded
as an extension.
\bigskip\bigskip
\centerline{\bf Acknowledgements}
\bigskip
\noindent
We would like to thank E. Bergshoeff, J. Dixon, M. Duff and K.S. Stelle
for useful discussions. E.S. would like to thank the International Center for
Theoretical Physics for hospitality.
\vfill\eject
\centerline{\bf APPENDIX}
\medskip
We collect here some well--known formulae on anomalies, which
have been used in the text. In particular we define the
cocycles $\omega^k_{p-k+2}$, for $k=0,1,2$, where $p-k+2$ is
the degree of $\omega$ as a form on $M$ (or on the world--volume,
if the connection is pulled back) and $k$ is
its degree as a form in the space of connections.
The Chern--Simons forms $\omega_{p+2}^0$ are defined
by the relation
$$
d\omega_{p+2}^0=k_p{\rm tr}F^{{p+3}\over2}\ ,\eqno(A.1)
$$
where $k_p$ is a normalization constant, depending on the group $G$,
which we will not specify (for the case $p=1$ see, for example, [12]).
For $p=1,3,5$ we have
$$
\eqalignno{
\omega_3^0(A) &= k_1{\rm tr} \bigg(FA -{1\over 3}A^3\bigg)\ ,     &(A.2a)\cr
\omega_5^0(A) &= k_3{\rm tr} \bigg(F^2A -\h F A^3+{1\over 10}A^5\bigg)\ ,
&(A.2b)\cr
\omega_7^0(A) &= k_5{\rm tr } \bigg(F^3A -{2\over 5}F^2A^3
-{1\over 5}FA FA^2 +{1\over 5}FA^5  -{1\over 35}A^7\bigg)\ . &(A.2c)\cr}
$$
The functional $C$ defined in (2.6) can be computed explicitly in the
cases $p=1,3,5$ by substituting (A.2) into (2.6). We get
$$
\eqalignno{
C_2=&\,k_1\tr(AL)\ ,&(A.3a)\cr
C_4=&{1\over4}k_3\tr\left[2(FA+AF-A^3)L+ ALAL-2AL^3\right]\ ,&(A.3b)\cr
C_6=&{1\over30}k_5\tr\bigl[
(10F^2A+10FAF+10AF^2-8FA^3-8A^3F-4AFA^2-4A^2FA+6A^5)L\cr
&+2F(A^2L^2-L^2A^2+3ALAL-3LALA)-6A^3LAL\cr
&+3F(LAL^2-L^2AL+2L^3A-2AL^3)+6A^3L^3\cr
&-3L^2A^2LA+3A^2L^2AL+2ALALAL+6L^3ALA+6AL^5\bigr]\ .
&(A.3c)\cr}
$$
The gauge variations of the Chern--Simons forms defines the
consistent anomaly $\omega^1_{p+1}$:
$$
\delta_\epsilon\omega^0_{p+2}(A)=d\omega^1_{p+1}(A,\epsilon)\ .\eqno(A.4)
$$
A general formula for $\omega^1_{p+1}$ can be found in [10].
It can be written in the form
$$
\omega^1_{p+1}(A,\epsilon)=\tr\,d\epsilon\,\phi_p(A)\ . \eqno(A.5)
$$
where the $p$-form $\phi_p=\phi_p^a T_a$ is a polynomial in $A$ and $F$.
For $p=1,3,5$ this polynomial is given by
$$
\eqalignno{
\phi_1=&-k_1 A\ ,&(A.6a)\cr
\phi_3=&-\ft12 k_3(FA+AF-A^3)\ ,&(A.6b)\cr
\phi_5=&-\ft13 k_5\left[(F^2A+FAF+AF^2)-\ft45(A^3F+FA^3)
-\ft25(A^2FA+AFA^2)+\ft35 A^5\right]\ . &(A.6c)\cr}
$$
Note that $\phi_p$ is also the coefficient of the term in $C$ linear in $L$.
The coboundary of $\omega^1_{p+1}$ defines $\omega^2_p$:
$$
\delta_{\epsilon_1}\omega^1(A,\epsilon_2)-
\delta_{\epsilon_2}\omega^1(A,\epsilon_1)-
\omega^1(A,[\epsilon_1,\epsilon_2])
=d\omega^2_p(A,\epsilon_1,\epsilon_2)\ .\eqno(A.7)
$$
For $p=1,3,5$ it is given by
$$
\eqalignno{
\omega^2_1(A,\epsilon_1,\epsilon_2)=&\ -2k_1\tr\,\epsilon_1 d\epsilon_2\ ,
&(A.8a)\cr
\omega^2_3(A,\epsilon_1,\epsilon_2)=&\ -k_3\tr\,\{d\epsilon_1,d\epsilon_2\}A\ ,
&(A.8b)\cr
\omega^2_5(A,\epsilon_1,\epsilon_2)=&\ {1\over15}k_5\tr\,
(5F-3A^2)\left[2A\{d\epsilon_1, d\epsilon_2\}-
d\epsilon_1 A d\epsilon_2+d\epsilon_2 A d\epsilon_1\right]\ .
&(A.8c)\cr}
$$
It is clear from (A.4) and (A.7) that $\omega^1_{p+1}$ and
$\omega^2_p$ are only defined up to a closed form.
In particular one could add to $\omega^1_{p+1}$
the closed form $-d(\tr\epsilon\phi(A))$ and get
$$
\hat\omega^1_{p+1}(A,\epsilon)=-\tr\,\epsilon d\phi_p\ ,\eqno(A.9)
$$
which is another form of the consistent anomaly. Applying the
coboundary to $\hat\omega_{p+1}^1$ defines a different 2-cocycle
$\hat\omega_p^2$:
$$
\delta_{\epsilon_1}\hat\omega^1(A,\epsilon_2)-
\delta_{\epsilon_2}\hat\omega^1(A,\epsilon_1)-
\hat\omega^1(A,[\epsilon_1,\epsilon_2])
=d\hat\omega^2_p(A,\epsilon_1,\epsilon_2)\ .\eqno(A.10)
$$
For $p=1,3,5$
$$
\eqalignno{
\hat\omega^2_1(A,\epsilon_1,\epsilon_2)=&\
k_1\tr\,[\epsilon_1,\epsilon_2]\,A\ ,
&(A.11a)\cr
\hat\omega^2_3(A,\epsilon_1,\epsilon_2)=&\
\ft12 k_3\tr\,\left([\epsilon_1,\epsilon_2](FA+AF-A^3)
-\epsilon_1 dA\epsilon_2 A-\epsilon_1 A\epsilon_2 dA\right)\ ,
&(A.11b)\cr
\hat\omega^2_5(A,\epsilon_1,\epsilon_2)=&\
\ft13 k_5\tr\big\{
[\epsilon_1,\epsilon_2]\left[(F^2A+FAF+AF^2)-\ft45\{A^3,F\}
-\ft25\{A,AFA\}+\ft35 A^5\right] \cr
&-\ft1{5}\,[\epsilon_1,d\epsilon_2][F,A^2]
-\ft35(d\epsilon_1 A\epsilon_2 +\epsilon_2 A d\epsilon_1)(FA+AF-A^3)\cr
&+\ft1{5}\,[\epsilon_2,d\epsilon_1][F,A^2]
-\ft35(d\epsilon_2 A\epsilon_1 +\epsilon_1 A
d\epsilon_2)(FA+AF-A^3)\big\}\ ,
&(A.11c)\cr}
$$
These are the cocycles one gets in the Gauss law algebra of an
anomalous fermionic theory using the Bjorken-Johnson-Low procedure
[13], or in the gauged Wess-Zumino-Witten model at the canonical level
[11]. They differ from
the cocycles $\omega^2_p$ by a redefinition of the current.

These cocycles assume a simpler form when their argument is $L$
instead of $A$. We have
$$
\eqalignno{
\omega^0_{p+2}(L)=&\,k_p c_p\tr\,L^{p+2}\ ,&(A.12a)\cr
\omega^1_{p+1}(L,\epsilon)=&\,k_p c_p(p+2)\tr\,d\epsilon\,L^p\ ,&(A.12b)\cr
\omega^2_p(L,\epsilon_1,\epsilon_2)=&\,
2k_p c_p(p+2)\tr\,\{d\epsilon_1,d\epsilon_2\}L^{p-2}\ ,&(A.12c)\cr}
$$
where $c_p=(-1)^{{p+1}\over2}
\left({{p+3}\over2}\right)\Gamma({{p+3}\over2})^2/\Gamma(p+3)$.

The quantity $\chi^a_{p-1}$ defined by (4.13) and (2.5) can be calculated
inserting (A.6) in (4.13). For the cases $p=1,3,5$ we find
$$
\eqalignno{
\chi^a_0 =&\ 0\ ,  &(A.13a) \cr
\chi^a_2 =&-\ft12k_3 {\rm tr}\ T^a[A,L]\ ,  &(A.13b)  \cr
\chi^a_4 =&\ k_5{\rm tr}\ T^a\big[-\ft16(\{F,[A,L]\}+AFL-LFA)
-\ft3{10}([ALA,A]+2[A^3,L])\cr
&+\ft1{10}([A^2,L^2]+3[ALA,L])
+\ft3{10}([LAL,L]+2[L^3,A])
\big]\ . &(A.13c) \cr}
$$

\bigskip
\centerline{\bf REFERENCES}
\medskip
\item{1.}
J.A. Dixon, M.J. Duff and E. Sezgin, Phys. Lett. {\bf 279B} (1992)
265.
\item{2.} J. Mickelsson, Lett. Math. Phys. {\bf 7} (1983) 45;
Comm. Math. Phys. {\bf 97} (1985) 361;\hfil\break
L. Faddeev, Phys. Lett. {\bf 145B} (1984) 81.
\item{3.} E. Bergshoeff, R. Percacci, E. Sezgin, K.S. Stelle and
P.K. Townsend, ``Realizations of gauge transformations in p--loop
space'', in preparation.
\item{4.} E. Bergshoeff, E. Sezgin and P.K. Townsend, Phys. Lett. {\bf 189B}
(1987) 75.
\item{5.} E. Witten, Comm. Math. Phys. {\bf 92} (1984) 455;\hfil\break
E. Bergshoeff, P.S. Howe, C.N. Pope, E. Sezgin and E. Sokatchev,
Nucl. Phys. {\bf B 354} (1991) 113.
\item{6.}
E. Bergshoeff, F. Delduc and E. Sokatchev, Phys. Lett. {\bf 262B}
(1991) 444.
\item{7.} A.C. Davis, J.A. Gracey and A.J. Macfarlane,
Phys. Lett. {\bf 194B} (1987) 415.
\item{8.} J.A. Dixon and M.J. Duff, preprint CTP-TAMU-45/92.
\item{9.} C.M. Hull and E. Witten, Phys. Lett. {\bf 160B} (1985) 798.
\item{10.}
J. Ma\~nes, R. Stora and B. Zumino, Comm. Math. Phys. {\bf 102} (1985)
157.
\item{11.} R. Percacci and R. Rajaraman, Phys. Lett. {\bf 201B} (1988)
256; Int. J. Mod. Phys. {\bf A 4} (1989) 4177.
\item{12.}
E. Bergshoeff, H. Sarmadi and E. Sezgin, in {\it Supersymmetry and its
Applications}, Eds. G.W. Gibbons, S.W. Hawking and P.K, Townsend
(Cambridge University Press, 1986).
\item{13.} S. Jo, Phys. Lett. {\bf 163B} (1985) 353.

\end